# THE NEED FOR EFFECTIVE INFORMATION SECURITY AWARENESS PRACTICES IN OMAN HIGHER EDUCATIONAL INSTITUTIONS

*Mr. Rajasekar Ramalingam (rajasekar.sur@cas.edu.om) \**
*Mr. Shimaz Khan (shimaz.sur@cas.edu.om) \**
*Mr. Shameer Mohammed (shameer.sur@cas.edu.om) \**

*\* Ministry of Higher Education, Sur College of Applied Sciences,
Department of Information Technology, Post Box: 484 Post Code: 411, Sultanate of Oman*

**ABSTRACT**

*The revolution of internet technology and its usage have led a significant increase in the number of online transactions and electronic data transfer, parallely increased the number of cybercrime incidents around the world. Steady economic growth in the Sultanate of Oman accelerated the volume of online utilization for e-commerce, banking, communication, education and so forth. Normally attackers target the users who ignore security practices due to the lack of information security awareness. Unawareness of information security practices, user negligence, lack of awareness programs and trainings are the root cause for information security threats. Earlier studies reveal there is a considerable and continuous cybercrime incident in Oman which compromises the security policy of the organizations, affecting the business continuity and the economic growth. In this study, a survey was performed among the educational institutions in Oman to investigate the level of information security awareness and based on the study; a security awareness model is proposed to enable information security practices in the educational institutions.*

## 1. Introduction

According to the word internet usage statistics news [2], Internet users in the Sultanate of Oman have been continuously increasing in recent days. Oman is noteworthy for marketers due to their excellent consumer protective laws, making online purchases safer and boost up the online purchasing [4]. Steady economic growth has increased the volume of the net banking and mobile banking. Improvements in payment infrastructure have driven the growth of Oman's card payments channel and it is forecasted the growth of card usage will be around 4.4 million by 2017 [5]. The high number of internet usage, increased card usage, and the revolution of internet technology has led a significant increase in the number of online transactions and electronic data transfer. Parallely, the number of cybercrime incidents in Oman is also increased. According to the annual report of Information Technology Authority (ITA) (2012 and 2013) [6], there is a significant increase in the number of cybercrime incidents in Oman. Compared to 2011, there was an increase of 13.5% reported incidents and 200% increase of malware incidents. 1,084,369 malicious attempts were prevented and analyzed against government portals. 19,171 malicious attempts against government networks were identified and prevented. 25,827 vulnerabilities were discovered by scanning 9,890 IPs. 941,079 malwares were analyzed to determine the main cause of infection. 659,090 web violations were analyzed and prevented. 15,855 security attacks were discovered and handled by OCERT. In the present ICT era, new threats are continuously evolving such as spam emails [1], identity theft [3], data leakage [7], phishing [8], adware [9], intrusion [10] and many more which would have considerable impact on the information assets of the organization and individuals. Attackers have adopted stealthier techniques to exploit user trust, this strategy leads to the most variable, unpredictable and critical vulnerability to the information assets. The gap between the knowledge and the practices regarding the security practices is the major factor which influences security of Information.

In this research paper a survey was performed among the education institutions in Oman to investigate the level of information security awareness among various entities which includes students, technical staff and academic staff by focusing on the factors like demographics, internet usage, awareness about organization's network, security threats experience, awareness on password management, email security awareness and the knowledge on security practices. The survey attracted 173 respondents; the level of information security awareness and knowledge on security practices were correlated and analyzed. The areas of weakness related to the security implementation, awareness, and practice were identified and a security model has been proposed.

## 2. Need for ISA in the higher educational institutions

Student records, employee records, payroll information, students grade information, research data, university policies, and intellectual property produced by research departments, colleges and universities house a vast amount of sensitive data and these are valuable IT assets to an institution. Websites, software and applications, computers, and network devices are also important assets of an institution. These assets have higher risk for data breaches and other IT security risks [12]. The study made by Bulgurcu, Cavusoglu & Benbasat [11], indicates that an employee's attitude and outcome beliefs are influenced by their personal level of ISA. Also it indicates that the ISA can positively influence the employee to comply with the security policies of an organization. Also the study made by the Hagen, Albrechtsen & Hovden [13] found the fact that the IT security measures can be successful only when there is emphasis on the ISA. These factors stress the importance of ISA in the successful implementation of IT security measures in any organization especially in the educational institutions.

## 3. Information Security Awareness Identification Model (ISAIM)

The proposed model, Information Security Awareness Identification Model (ISAIM) has been adopted in this study as shown in Figure 1.1. ISAIM has 6 key elements which include: effective usage, organization awareness, threats awareness, protection awareness, content awareness and security practice. Effective usage element identifies how frequent the user access IT enabled devices, the purpose of usage and access location. Organization awareness element identifies the user's knowledge about their environment such as availability of IT infrastructure, security policy, and security standards. Threats awareness element refers to finding the threat experience of an individual, the frequency, loss due to attacks, knowledge of security policy and reporting mechanism. Protection awareness element identifies how efficiently identities of the individuals are chosen and managed. Content awareness element identifies how the user evaluates the validity of email content. Security practice element identifies the necessary components of security habits such as skills needed, training attended and training required.

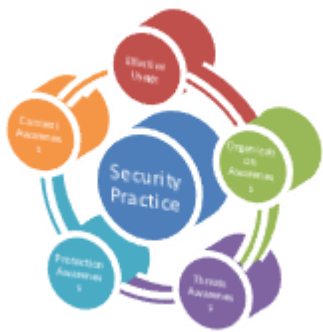

**Figure 1.1: ISAIM Model**

**4. Survey target group**

The target group is the entities of the educational institutions in Oman which includes the academic staff, technical staff and students. The reasons for choosing the group are, academic staffs are the main pillars of educational institutions; they are playing a vital role in articulating the technical knowledge of the students and major user of Information Technology. Indeed, it is essential to find the level of security awareness that exist among the academic staff community. The purpose of finding the level of security awareness among the technical staff is to find how the technology is used to protect the information assets of any institution by them. Students are our ambassadors and they are going to implement all the skills and knowledge learned in the various organizations of Oman in the near future. They must know how to secure their work environment and the information they are dealing with. They do require a detailed understanding of the security threats, damages by various threats and solutions to mitigate the damage. Table 1.1 shows the list of educational institutions participated in the survey.

**Table 1.1 List of Oman educational institutions participated in survey**

| S# | Name of the Educational Institution | S# | Name of the Educational Institution |
|---|---|---|---|
| 1 | Al Buraimi University College | 10 | Sohar College of Applied Sciences |
| 2 | Higher College of Technology | 11 | Nizwa College of Technology |
| 3 | Ibra College of Technology | 12 | Oman College of Management Technology |
| 4 | Salalah College of Technology | 13 | Al Sharqiyah University |
| 5 | Sur College of Applied Sciences | 14 | German University of Technology in Oman |
| 6 | Waljat College of Applied Sciences | 15 | Ibri College of Applied Sciences |
| 7 | Majan University College | 16 | Sultan Qaboos University |
| 8 | College of Applied Sciences, Rustaq | 17 | Caledonian College of Engineering |
| 9 | Sohar University | | |

## 5. ISAIM implementation

This section describes the implementation of ISAIM methodology to find the level of Information security awareness. The survey response from the various respondents were analyzed, the key findings are as follows:

### 5.1 Effective Usage

Among 173 respondents, 76.9% were male and 21.3% were female, most of the respondents fall into the age category of 18 to 29. The educational qualifications of the respondents are categorized as, 35% graduates, 38% masters and 23% PhD holders. The numbers of the academic staffs were 54% and the reset were students and technical staff. The majority of the respondents (70%) use smart phones or mobile network to access internet. As far as the location of the internet usage is concerned, majorities were accessing the internet from home, second majority were accessing from the educational institutions and only 7% were accessing from the internet cafes. Emailing and educational references are the major purpose for using the internet where as social networking and net banking holds the next highest level of internet usage.

### 5.2 Organizational awareness

Surprisingly, 22% of institutions are using any Information Security practices and 39% of the respondents were unaware about the security practices in use. The antivirus software and firewall are the widely used information security tools at the educational institutions of Oman. Most of the educational institutions use firewall as the security tool to safe guard their network. It has been revealed that the software's tools like Intrusion Detection System (IDS) and DMZ are still a mysterious term to the respondents. Most of the respondents doesn't have the knowledge whether the IDS is deployed or not. 61% of the respondents don't know about DMZ software tool. Very few respondents revealed that "I don't use any security software".

### 5.3 Threat Awareness

On an average, it has been found from the survey that 71% of the respondents has encountered with security attacks in a range of 1 to 3 times. Only 13% respondents have become victims of security attacks more than 10 times per year. From the security attacks, 36% of the respondents have lost their personal data and 35% of the respondents systems got completely crashed. Virus, spam emails and adware are the three major causes for the security attacks. 39% respondents revealed that they are not aware of the security policy of the institution and 70% of the institutions are not having any security practices. 38% of the respondents are not aware of the reporting mechanisms, 25% of institutions do not have any reporting mechanism and most of the respondents are not interested in reporting the security incidents itself.

### 5.4 Password Awareness

The survey focused on the password management also and found that most of the respondents are well aware of password management strategies. 50% of the respondents are maintaining different passwords for different applications, few of the respondents, around 17% maintains common password for different applications. 50% of the respondents make use of alphabets, numbers, and special characters to generate the password and 59% of respondents prefer to maintain the password length of 6-9 characters long. Survey reveals that 84% of respondents do memorize their passwords and 78% are not ready to share their passwords across the website, where as 7% are ready for the same. 19% of the respondents are not at all changing their passwords and 57% of the respondents are changing their passwords provided the application which they use will insist to change their passwords.

### 5.5 Content Awareness

32% of the respondents showed their concern about opening emails from unknown sources. 39% of the respondents ratified that there is no email policy in their institution and 23% of the respondents revealed that they are aware of the policies, but they conveyed as "I do not know and I could not understand". Phishing emails gather the user's personal information; 84% of the respondents conveyed that they don't share their personal information and very meagre number of respondents i.e. 3% is willing to provide their banking credentials through emails. 79% of the respondents are not willing to encourage responding to hoax and chain mails.

### 5.6 Security Practice

The majority of respondents, i.e. 64% consider that the proper knowledge, good skills, and clean attitude are the vital components for developing a positive security practices. There are nearly 40% of the respondents are not self-assured regarding their institution protection against the Information security risk. 56% of the respondents conveyed that their institution has never conducted any security awareness program and 59% of the respondents confirmed that they have not attended any training programs during the last 12 months.

## 6. Conclusion

Information security awareness is an essential and foundational element in protecting and assuring the nation's information assets and always ready to meet mission objectives. The study on Information security awareness in educational institutions found several important issues that need to be addressed. Although the student, technical staff and academic staff are having the basic knowledge of security, still they are not aligned to the security practice of the institutions. There seems to be a significant lack of understanding in protecting information assets of the institutions. There needs to be a greater sense of urgency on the part of the government, other professional bodies and the educational institution to educate users about the information security needs of an institution. Implementing awareness training programs will never solve the purpose. The educational institutions must be tested and held accountable to ensure information security policies and procedures are understood and followed by their entities. This survey identified that, as an individual, the knowledge of information security awareness is considerably better but as an institution, information security awareness should be improved and there is an immediate need for security standards, policy, reporting mechanism and continuous awareness training in the educational institutions of Oman. Information security awareness is not get-it and forget-it activity, it needs periodical support and propose an awareness program specific to educational institutions in Oman.